\documentclass[twocolumn,twoside,slac_two, tightenlines, nofootinbib,natbib]{revtex4}

\usepackage{graphicx}
\usepackage{fancyhdr}
\usepackage{aas_macros}
\usepackage{lineno}
\usepackage{graphicx}
\usepackage{amssymb}
\usepackage{amsmath}
\usepackage{times}
\usepackage{courier}
\usepackage{xspace}

\newcommand{\gr}{$\gamma$-ray\xspace}
\newcommand{\grs}{$\gamma$ rays\xspace}

\newcommand{\beq}{\begin{equation}}
\newcommand{\eeq}{\end{equation}}

\newcommand{\sigv}{\ensuremath{\langle\sigma v\rangle}\xspace}
 
\renewcommand{\deg}{\ensuremath{^{\circ}}\xspace}
\newcommand{\gev}{\xspace\ensuremath{\mathrm{GeV}}\xspace}

\newcommand{\mev}{\xspace\ensuremath{\mathrm{MeV}}\xspace}

\newcommand{\zetap}{\ensuremath{\zeta_{p,\mathrm{max}}}\xspace}
\begin{document}

\title{Galaxy Clusters with the Fermi-LAT: Status and Implications for Cosmic Rays and Dark Matter Physics}

%

\author{S. Zimmer}
\affiliation{Oskar Klein Center for Cosmoparticle Physics and Department of Physics, Stockholm University, Stockholm, SE 10691, Sweden}
\author{for the \textit{Fermi}-LAT Collaboration}
\noaffiliation

\begin{abstract}
Galaxy clusters are the most massive systems in the known universe. They host relativistic cosmic ray populations and are thought to be gravitationally bound by large amounts of Dark Matter, which under the right conditions could yield a detectable \gr\, flux. Prior to the launch of the Fermi satellite, predictions were optimistic that Galaxy clusters would be established as \gr-bright objects by observations through its prime instrument, the Large Area Telescope (LAT). Yet, despite numerous efforts, even a single cluster detection is still pending.
\end{abstract}

\maketitle

\thispagestyle{fancy}

\section{Introduction}

Galaxy clusters (GC) represent the largest virialized objects that are believed to have formed through a hierarchical build up of structures over the evolution of the universe. In this picture, baryonic matter accretes towards the gravitational well caused by large amounts of Dark Matter (DM) which make up 26\% of the energy density of the Universe \citep{Ade:2014aa}. Through gravitational interaction, smaller structures merge with one another, forming groups of galaxies and eventually clusters. Determining the exact nature of DM is one of the greatest challenges of modern physics and weakly interactive massive particles (WIMPs) prove to be strong candidates fulfilling the role as DM particle \citep{Bergstrom:1999aa,Bergstrom:2009aa}. The neutralino, which in several extensions of the standard model of particle physics is predicted to be the lightest stable supersymmetric particle, provides a natural WIMP candidate. In many of these models the neutralino can self-decay or annihilate into lighter standard model particles, among others high energy \grs which, if observed, can be used to trace back the origin of the interaction and indirectly detect DM \citep{Baltz:2008aa}. 

While clusters are promising targets due to their large DM content, predicted \gr emission on top of that of individual cluster member galaxies constitutes an irreducible foreground. This foreground emission arises from cosmic ray (CR) interactions with the intra-cluster medium (ICM) and is motivated by conventional astrophysics \citep[see, e.g.][for a review]{Petrosian:2008aa}, while the observation of DM-induced \grs may be regarded as a somewhat more exotic signature \citep{Pinzke:2011aa}. 

In this contribution I will review the most recent studies of GCs with the Fermi Large Area Telescope (LAT) undertaken by the instrument team. I will start by summarizing recent efforts aiming at the astrophysical emission scenario of CR interactions resulting in a detectable \gr flux (Section~\ref{sec:cr}) and briefly report on work in progress in regards to DM constraints that can be obtained from GCs (Section~\ref{sec:dm}). Finally, I will discuss one of the challenges involved when searching for large extended sources such as GCs (Section~\ref{sec:virgo}) and conclude by commenting on the implications for future searches with the LAT. 

\section{Cosmic Ray Induced \grs\label{sec:cr}}

The majority of the baryonic mass in GCs is present in the form of hot ionized gas, the ICM, which has been detected via thermal X-ray emission observed by contemporary space telescopes such as ROSAT or XMM-Newton \citep[see, e.g.][for a review]{Kaastra:2008aa}. In addition, large scale radio synchrotron emission has been detected in a number of the most nearby clusters which can be classified into halos and relics \citep[][]{Ferrari:2008aa}. The latter appear polarized, while the former are not, suggesting a different emission mechanism to be at play. The observation of radio-synchrotron emission indicates the presence of a pool of relativistic electrons (CRe). Together with magnetic fields this provides a favorable environment for high energy particle interactions between the CRes and the ICM which may be observable through the detection of \grs or hard X-rays \citep[see, e.g. the excellent review by][]{Brunetti:2014aa}. However, due to the short diffusion times, CRes must be constantly replenished, e.g. through injection by active galactic nuclei (AGN) or be created through secondary processes. 

Another intriguing possibility are hadronically-induced \grs. Here CR protons (CRp) may be accelerated within the ICM through means of diffusive shock acceleration (DSA) and due to their large diffusion time remain within the cluster volume. CRp then interact with the ICM and produce \grs via decay of neutral pions. The latter has received particular attention as \citet{Pinzke:2010aa} have shown the emergence of a universal model describing the CR interactions in a cosmological framework based on smooth-particle hydrodynamics simulations. In their model, the resultant \gr spectrum is dominated by the aforementioned $\pi^{0}$ decay and IC emission is essentially negligible. The spectrum shows the characteristic $\pi^{0}$ bump at $\sim130~\mev$ and for energies $>500~\mev$ follows a powerlaw with index $2.3$.\footnote{The interested reader is referred to \citet{Pinzke:2010aa}. The true spectrum is concavely shaped but for the considered LAT energies, it can approximated with a powerlaw as discussed in the main text.} The resulting spatial distribution is close to that of the thermal X-rays as it is expected that the CRs are following the gas. One key assumption when creating the model is that the maximum injection efficiency, \zetap at which protons can be accelerated via DSA is similar to that what has previously been observed in SNRs \citep{Helder:2009aa}.\footnote{In their works, the authors adopt $\zetap=50\%$ as benchmark when calculating their \gr predictions \citep{Pinzke:2011aa}.} Together with the claimed universality of the spectrum, this allowed us to employ the joint likelihood technique, a statistical method in which each target is optimized according to its individual nuisance parameters and then the information from each individual likelihood is combined into a global likelihood by multiplying them \citep[see][for a technical discussion and applications]{Anderson:2014aa}. 

The starting point for the study in \citet{Ackermann:2014ab} has been the extended HIFLUGCS catalog, a X-ray flux-limited complete sample of nearby GCs. Selecting a set of 50 clusters, we found a global excess at the level of $\sim2.7\sigma$ which however could be entirely attributed to previously non-detected individual cluster member galaxies (with known  counterparts in the radio band). Thus, with four years of LAT data, no cluster was detected and flux upper limits were set. The most constraining cluster in the sample is the Coma cluster with a reported integral flux limit of $4.0\times10^{-10}\mathrm{ph/cm^{2}/s}$ assuming an extended emission characteristic according to the benchmark model by \citet{Pinzke:2010aa} and $2.5\times10^{-10}\mathrm{ph/cm^{2}/s}$ when considering a pointlike emission.\footnote{These limits were calculated over the entire energy range of 500~\mev to 200~\gev.} Based on the joint likelihood approach, we also find that in order to account for the non-observation, DSA must be either substantially less efficient ($\zetap\lesssim21\%$) or conversely, the CR-to-thermal pressure ratio must be lower than $1\%$, making the contribution of CRp's to the ambient \gr flux negligible \citep[see also the discussion in][]{Vazza:2014aa}.

\section{Dark Matter constraints from Cluster Observations\label{sec:dm}}
Given its non-detection, ongoing searches for \grs from GC can thus be used to constrain the available parameter space of WIMP DM. Generically, the induced \gr\ flux from WIMP pair annihilation can be expressed as 
\beq
\phi_{s}(\Delta\Omega)=\underbrace{\frac{1}{4\pi}\frac{\sigv}{m^{2}_{DM}}}_{\Phi_{PP}}\times\underbrace{\int_{\Delta\Omega}\int_{l.o.s.}\rho^{2}(\emph{r})dl~d\Omega'}_{J-factor}.
\eeq
In the above equation $\Phi_{PP}$ refers to the particle physics term containing both the mass of the WIMP and its velocity-averaged pair annihilation cross section \sigv. The second term is referred to as astrophysical, or $J$-factor and is the line of sight integral of the DM column density. N-body simulations suggest that DM clusters (clumps) across all mass scales, forming sub haloes in addition to the smooth main halo. The amount of substructure as well as the properties are largely unknown and current N-body simulations do not have the capabilities yet to resolve the smallest substructures. Hence, extrapolations over several orders of magnitude are necessary. For this (abbreviated) discussion it is sufficient to address the amount of substructure by the introduction of a dimensionless boost factor $b$, which relates the $J$-factor obtained by assuming a universal NFW halo \citep{Navarro:1997aa} to that obtained when considering different amounts of substructure. Depending on the extrapolation scheme, boost factors for clusters may vary between O(30) to O(1000) \citep[see e.g. the discussion in][]{Sanchez-Conde:2011aa}. While the predicted annihilation flux profile is similar for both model predictions, the overall predicted flux may vary by orders of magnitude. This fact makes DM constraints from clusters (at least as far as annihilation is considered) more model dependent than e.g. those obtained from the observation of nearby dwarf spheroidal galaxies \citep{Ackermann:2014aa}.\footnote{Note that the question of substructure is typically less important in the case of decay as the associated $J$-factor scales linearly with the DM density.} 

\begin{figure*}
\includegraphics[width=\columnwidth]{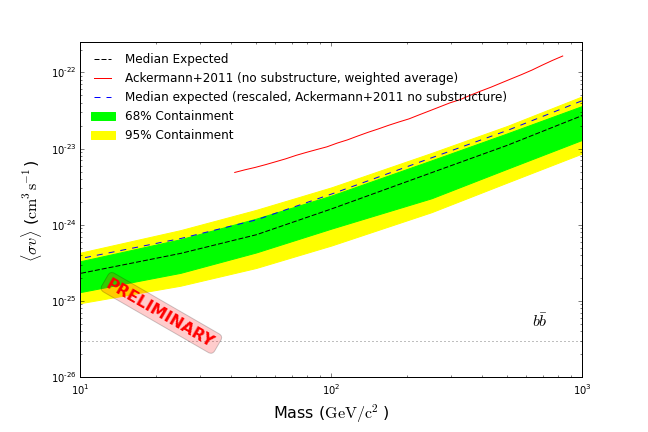}
\includegraphics[width=\columnwidth]{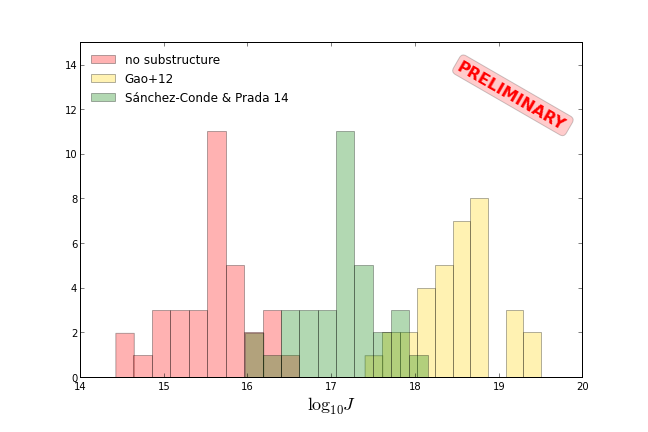}
\caption{\label{fig:dmLimits}\emph{Left:} Expected sensitivity (evaluated by calculating the combined one-sided 95\% C.L. upper limit) for annihilating WIMP DM into $b\overline{b}$ from a combined analysis of 34 galaxy clusters in 5 years of LAT data using photons from 500\mev-500\gev. Each cluster model contains a conservative amount of substructure. The red solid line corresponds to the first LAT team publication based on 11 months of data \citep{Ackermann:2010aa} in which each individual limit has been weighted with the assumed J-factor (NFW only). \emph{Right:} Distribution of $J$-factors for our chosen cluster sample. In the main text we refer to the setup by \citet{Sanchez-Conde:2014aa} as \emph{conservative} setup. The \emph{optimistic} configuration refers to the distribution labeled with \citet{Gao:2012aa}.} 
\end{figure*}

Ongoing work (based on a five year dataset) is focusing yet again on a subsample of the most massive nearby clusters, selected from the X-ray flux limited HIFLUCGS sample. We demand there to be no appreciable overlap by requiring a distance between each cluster of the sum of the virial radius of each cluster along with a 2\deg buffer accounting for the tails in the LAT point spread function (PSF). For the resulting 34 clusters we construct spatial templates according to substructure models considering both a \emph{conservative} boost factor of O(30) \citep{Sanchez-Conde:2014aa} and contrast this with a more \emph{optimistic} choice of O(1000) \citep{Gao:2012aa}. For both configurations we perform a binned likelihood analysis. After having found the best fit parameters of our background fit, we construct a bin-by-bin likelihood function by assuming a simple powerlaw with index 2.0 in each energy bin which would account for the cluster emission. The advantage is that the resulting flux limits can be used to directly test spectrally different models without the need of repeating the entire likelihood procedure \citep[see][for details regarding the bin-by-bin method]{Ackermann:2014aa}. In Fig.~\ref{fig:dmLimits} we show both the J-factor distribution from our chosen sample as well as the estimated sensitivity by selecting high-latitude regions in the sky which are selected such that the center a) does not contain a 3FGL source \citep{The-Fermi-LAT-Collaboration:2015aa} and b) does not coincide with a cluster center or a circular region with a radius corresponding to the virial radius of the cluster.\footnote{For this analysis we select the subset of P7REP photons that pass the CLEAN class and apply the recommended instrument response function {\tt P7REP\_CLEAN\_V15}.}

\section{Challenging individual targets: very extended emission from the Virgo cluster\label{sec:virgo}}

While the discussed emission mechanisms vary appreciably with regards to the spectral form of the predicted emission, the studies that I discussed here have in common that the targets are large extended sources.\footnote{With large we refer to an emission radius of $\sim2-3\deg$ (as in the case of the Fornax and Coma cluster, respectively).} However, even among these extended sources, there are extreme cases. The largest target is the Virgo cluster, our closest neighbor which appears as a structure in the Northern part of the sky that extends up to 14\deg in diameter. The cluster itself is undergoing a complex merger in which the main clumps centered around the giant ellipticals M87 and M49 are moving towards each other. 

\begin{figure}
\begin{center}
\includegraphics[width=\columnwidth]{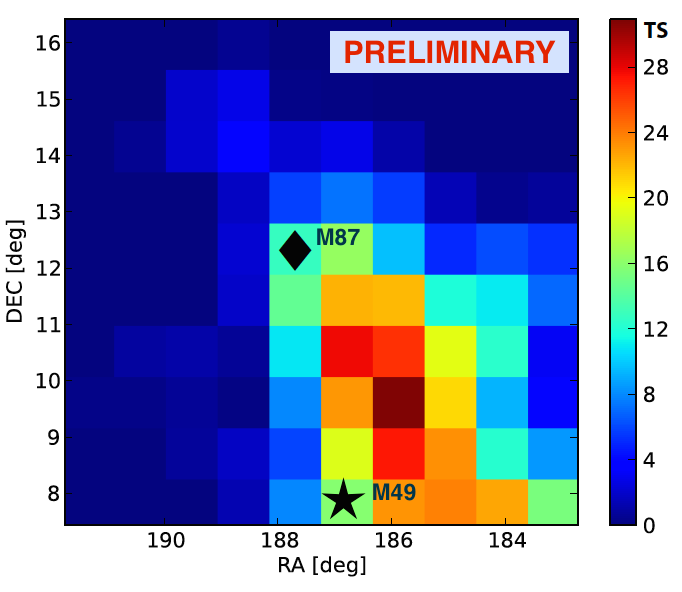}
\end{center}
\caption{\label{fig:VirgoTS}Grid scan of TS values when placing a uniform disk emission template with a powerlaw and $\gamma=2.0$ on various positions in the map. This representation allows to map the observed excess when the foreground Galactic diffuse emission is modeled using the standard template for point source analysis. For reference we add the reported NED positions for the central galaxies M87 and M49 as diamond and star-shaped markers, respectively.}
\end{figure}

The poor PSF at low energy together with the large surface area require special care when searching for an extended emission contribution, as recently claimed \citep{Han:2012aa,Macias-Ramirez:2012aa}. It is important to emphasize that the model for the Galactic foreground emission that is usually employed when analyzing Fermi-LAT data is optimized for point source searches. Indeed, when confronting a dataset comprising three years of collected photons between 100\mev and 100\gev, we find an extended excess if we employ the standard diffuse model \citep{Ackermann:2015aa}.\footnote{For this analysis we select {\tt Pass 7} (P7V6) photons passing the {\tt SOURCE} selection together and apply the recommended models for modeling the Galactic diffuse and isotropic emission. The reader is referred to the web pages of the Fermi Science Support Center for details: \url{http://fermi.gsfc.nasa.gov/ssc/data/access/lat/BackgroundModels.html}} However, when systematically performing a scan over the entire ROI by using a uniform disk of 3\deg radius, we find two distinct maxima which are spread out and appear away from both sub clump centers as shown in Fig.~\ref{fig:VirgoTS}. Moreover, when using a set of alternatively derived diffuse foreground models \citep{Ackermann:2012aa,de-Palma:2013aa}, the significance of this extended excess varies appreciably, implying that the source of the excess may be due to an incomplete modeling of the Galactic foreground emission.

\section{Outlook}
Despite intense efforts no \gr emission has been detected from clusters to date. In the meantime radio observations have revealed more and more systems containing extended, yet faint radio sources. This radio emission will remain the driving force when searching for non-thermal emission from galaxy clusters. The LAT with its continuous sky survey capabilities will remain instrumental in probing the important $\sim\mev-\gev$ domain which is much too low for air Cherenkov telescopes to be sensitive to. In particular the extension towards lower energies, enabled by the latest reconstruction algorithms, collectively dubbed \emph{Pass 8} will help in achieving this goal \citep{Atwood:2013aa}, as it provides a better PSF and an increased acceptance towards lower energies. 

As for DM constraints, clusters will remain challenging targets - due to their large extension and the intrinsic uncertainties in their $J$-factors. However, they are also complementary targets to probe in case evidence arises from more promising targets such as dwarf galaxies or the Galactic center. 

\begin{acknowledgments}
The \textit{Fermi}-LAT Collaboration acknowledges support for LAT development, operation and data analysis from NASA and DOE (United States), CEA/Irfu and IN2P3/CNRS (France), ASI and INFN (Italy), MEXT, KEK, and JAXA (Japan), and the K.A.~Wallenberg Foundation, the Swedish Research Council and the National Space Board (Sweden). Science analysis support in the operations phase from INAF (Italy) and CNES (France) is also gratefully acknowledged.

I wish to acknowledge the financial travel support of the W.~Kobbs and Grundstr\"{o}ms foundation and thank the organizers for a stimulating meeting. 

\end{acknowledgments}


\bibliography{proc}

\begin{thebibliography}{27}
\expandafter\ifx\csname natexlab\endcsname\relax\def\natexlab#1{#1}\fi
\expandafter\ifx\csname bibnamefont\endcsname\relax
  \def\bibnamefont#1{#1}\fi
\expandafter\ifx\csname bibfnamefont\endcsname\relax
  \def\bibfnamefont#1{#1}\fi
\expandafter\ifx\csname citenamefont\endcsname\relax
  \def\citenamefont#1{#1}\fi
\expandafter\ifx\csname url\endcsname\relax
  \def\url#1{\texttt{#1}}\fi
\expandafter\ifx\csname urlprefix\endcsname\relax\def\urlprefix{URL }\fi
\providecommand{\bibinfo}[2]{#2}
\providecommand{\eprint}[2][]{\url{#2}}

\bibitem[{\citenamefont{{Ackermann}}
  \emph{et~al.}(2014{\natexlab{a}})\citenamefont{{Ackermann}, {Ajello},
  {Albert}} \emph{et~al.}}]{Ackermann:2014ab}
\bibinfo{author}{\bibnamefont{{Ackermann}}, \bibfnamefont{M.}},
  \bibinfo{author}{\bibfnamefont{M.}~\bibnamefont{{Ajello}}},
  \bibinfo{author}{\bibfnamefont{A.}~\bibnamefont{{Albert}}}, \emph{et~al.},
  \bibinfo{year}{2014}{\natexlab{a}}, \bibinfo{journal}{\apj}
  \textbf{\bibinfo{volume}{787}}, \bibinfo{eid}{18}.

\bibitem[{\citenamefont{{Ackermann}}
  \emph{et~al.}(2010)\citenamefont{{Ackermann}, {Ajello}, {Allafort}}
  \emph{et~al.}}]{Ackermann:2010aa}
\bibinfo{author}{\bibnamefont{{Ackermann}}, \bibfnamefont{M.}},
  \bibinfo{author}{\bibfnamefont{M.}~\bibnamefont{{Ajello}}},
  \bibinfo{author}{\bibfnamefont{A.}~\bibnamefont{{Allafort}}}, \emph{et~al.},
  \bibinfo{year}{2010}, \bibinfo{journal}{\jcap} \textbf{\bibinfo{volume}{5}},
  \bibinfo{eid}{025}.

\bibitem[{\citenamefont{{Ackermann}}
  \emph{et~al.}(2012)\citenamefont{{Ackermann}, {Ajello}, {Atwood}}
  \emph{et~al.}}]{Ackermann:2012aa}
\bibinfo{author}{\bibnamefont{{Ackermann}}, \bibfnamefont{M.}},
  \bibinfo{author}{\bibfnamefont{M.}~\bibnamefont{{Ajello}}},
  \bibinfo{author}{\bibfnamefont{W.~B.} \bibnamefont{{Atwood}}}, \emph{et~al.},
  \bibinfo{year}{2012}, \bibinfo{journal}{\apj} \textbf{\bibinfo{volume}{750}},
  \bibinfo{eid}{3}.

\bibitem[{\citenamefont{{Ackermann}}
  \emph{et~al.}(2014{\natexlab{b}})\citenamefont{{Ackermann}, {Albert},
  {Anderson}} \emph{et~al.}}]{Ackermann:2014aa}
\bibinfo{author}{\bibnamefont{{Ackermann}}, \bibfnamefont{M.}},
  \bibinfo{author}{\bibfnamefont{A.}~\bibnamefont{{Albert}}},
  \bibinfo{author}{\bibfnamefont{B.}~\bibnamefont{{Anderson}}}, \emph{et~al.},
  \bibinfo{year}{2014}{\natexlab{b}}, \bibinfo{journal}{\prd}
  \textbf{\bibinfo{volume}{89}}(\bibinfo{number}{4}), \bibinfo{eid}{042001}.

\bibitem[{{Ackermann} \emph{et~al.}(2015)\citenamefont{{Ackermann}}
  \emph{et~al.}}]{Ackermann:2015aa}
\bibinfo{author}{\bibnamefont{{Ackermann}}, \bibfnamefont{M.}}, \emph{et~al.},
  \bibinfo{year}{2015}, \bibinfo{journal}{in preparation} .

\bibitem[{\citenamefont{{Ade}} \emph{et~al.}(2014)\citenamefont{{Ade},
  {Aghanim}, {Armitage-Caplan}} \emph{et~al.}}]{Ade:2014aa}
\bibinfo{author}{\bibnamefont{{Ade}}, \bibfnamefont{P.~A.~R.}},
  \bibinfo{author}{\bibfnamefont{N.}~\bibnamefont{{Aghanim}}},
  \bibinfo{author}{\bibfnamefont{C.}~\bibnamefont{{Armitage-Caplan}}},
  \emph{et~al.}, \bibinfo{year}{2014}, \bibinfo{journal}{\aap}
  \textbf{\bibinfo{volume}{571}}, \bibinfo{eid}{A16}.

\bibitem[{\citenamefont{{Anderson}}(2014, this conference)}]{Anderson:2014aa}
\bibinfo{author}{\bibnamefont{{Anderson}}, \bibfnamefont{B.}},
  \bibinfo{year}{2014, this conference}, in \emph{\bibinfo{booktitle}{eConf
  C141020.1}}.

\bibitem[{\citenamefont{{Atwood}} \emph{et~al.}(2013)\citenamefont{{Atwood},
  {Albert}, {Baldini}} \emph{et~al.}}]{Atwood:2013aa}
\bibinfo{author}{\bibnamefont{{Atwood}}, \bibfnamefont{W.}},
  \bibinfo{author}{\bibfnamefont{A.}~\bibnamefont{{Albert}}},
  \bibinfo{author}{\bibfnamefont{L.}~\bibnamefont{{Baldini}}}, \emph{et~al.},
  \bibinfo{year}{2013}, \bibinfo{journal}{ArXiv e-prints} \eprint{1303.3514}.

\bibitem[{\citenamefont{Baltz} \emph{et~al.}(2008)\citenamefont{Baltz, Berenji,
  Bertone} \emph{et~al.}}]{Baltz:2008aa}
\bibinfo{author}{\bibnamefont{Baltz}, \bibfnamefont{E.}},
  \bibinfo{author}{\bibfnamefont{B.}~\bibnamefont{Berenji}},
  \bibinfo{author}{\bibfnamefont{G.}~\bibnamefont{Bertone}}, \emph{et~al.},
  \bibinfo{year}{2008}, \bibinfo{journal}{JCAP}
  \textbf{\bibinfo{volume}{0807}}, \bibinfo{pages}{013}.

\bibitem[{\citenamefont{Bergstrom}(1999)}]{Bergstrom:1999aa}
\bibinfo{author}{\bibnamefont{Bergstrom}, \bibfnamefont{L.}},
  \bibinfo{year}{1999}, \bibinfo{journal}{Nucl. Phys. Proc. Suppl.}
  \textbf{\bibinfo{volume}{70}}, \bibinfo{pages}{31}.

\bibitem[{\citenamefont{Bergstrom}(2009)}]{Bergstrom:2009aa}
\bibinfo{author}{\bibnamefont{Bergstrom}, \bibfnamefont{L.}},
  \bibinfo{year}{2009}, \bibinfo{journal}{New J. Phys.}
  \textbf{\bibinfo{volume}{11}}, \bibinfo{pages}{105006}.

\bibitem[{\citenamefont{{Brunetti} and {Jones}}(2014)}]{Brunetti:2014aa}
\bibinfo{author}{\bibnamefont{{Brunetti}}, \bibfnamefont{G.}}, and
  \bibinfo{author}{\bibfnamefont{T.~W.} \bibnamefont{{Jones}}},
  \bibinfo{year}{2014}, \bibinfo{journal}{International Journal of Modern
  Physics D} \textbf{\bibinfo{volume}{23}}, \bibinfo{eid}{1430007}.

\bibitem[{\citenamefont{{de Palma}} \emph{et~al.}(2013)\citenamefont{{de
  Palma}, {Brandt}, {Johannesson}, {Tibaldo}, and {for the Fermi LAT
  collaboration}}}]{de-Palma:2013aa}
\bibinfo{author}{\bibnamefont{{de Palma}}, \bibfnamefont{F.}},
  \bibinfo{author}{\bibfnamefont{T.~J.} \bibnamefont{{Brandt}}},
  \bibinfo{author}{\bibfnamefont{G.}~\bibnamefont{{Johannesson}}},
  \bibinfo{author}{\bibfnamefont{L.}~\bibnamefont{{Tibaldo}}}, and
  \bibinfo{author}{\bibnamefont{{for the Fermi LAT collaboration}}},
  \bibinfo{year}{2013}, \bibinfo{journal}{ArXiv e-prints} \eprint{1304.1395}.

\bibitem[{\citenamefont{{Ferrari}} \emph{et~al.}(2008)\citenamefont{{Ferrari},
  {Govoni}, {Schindler}} \emph{et~al.}}]{Ferrari:2008aa}
\bibinfo{author}{\bibnamefont{{Ferrari}}, \bibfnamefont{C.}},
  \bibinfo{author}{\bibfnamefont{F.}~\bibnamefont{{Govoni}}},
  \bibinfo{author}{\bibfnamefont{S.}~\bibnamefont{{Schindler}}}, \emph{et~al.},
  \bibinfo{year}{2008}, \bibinfo{journal}{\ssr} \textbf{\bibinfo{volume}{134}},
  \bibinfo{pages}{93}.

\bibitem[{\citenamefont{Gao} \emph{et~al.}(2012)\citenamefont{Gao, Navarro,
  Frenk} \emph{et~al.}}]{Gao:2012aa}
\bibinfo{author}{\bibnamefont{Gao}, \bibfnamefont{L.}},
  \bibinfo{author}{\bibfnamefont{J.~F.} \bibnamefont{Navarro}},
  \bibinfo{author}{\bibfnamefont{C.~S.} \bibnamefont{Frenk}}, \emph{et~al.},
  \bibinfo{year}{2012}, \bibinfo{journal}{\mnras}
  \textbf{\bibinfo{volume}{425}}, \bibinfo{pages}{2169}.

\bibitem[{\citenamefont{{Han}} \emph{et~al.}(2012)\citenamefont{{Han}, {Frenk},
  {Eke}} \emph{et~al.}}]{Han:2012aa}
\bibinfo{author}{\bibnamefont{{Han}}, \bibfnamefont{J.}},
  \bibinfo{author}{\bibfnamefont{C.~S.} \bibnamefont{{Frenk}}},
  \bibinfo{author}{\bibfnamefont{V.~R.} \bibnamefont{{Eke}}}, \emph{et~al.},
  \bibinfo{year}{2012}, \bibinfo{journal}{\mnras}
  \textbf{\bibinfo{volume}{427}}, \bibinfo{pages}{1651}.

\bibitem[{\citenamefont{{Helder}} \emph{et~al.}(2009)\citenamefont{{Helder},
  {Vink}, {Bassa}} \emph{et~al.}}]{Helder:2009aa}
\bibinfo{author}{\bibnamefont{{Helder}}, \bibfnamefont{E.~A.}},
  \bibinfo{author}{\bibfnamefont{J.}~\bibnamefont{{Vink}}},
  \bibinfo{author}{\bibfnamefont{C.~G.} \bibnamefont{{Bassa}}}, \emph{et~al.},
  \bibinfo{year}{2009}, \bibinfo{journal}{Science}
  \textbf{\bibinfo{volume}{325}}, \bibinfo{pages}{719}.

\bibitem[{\citenamefont{{Kaastra}} \emph{et~al.}(2008)\citenamefont{{Kaastra},
  {Paerels}, {Durret}} \emph{et~al.}}]{Kaastra:2008aa}
\bibinfo{author}{\bibnamefont{{Kaastra}}, \bibfnamefont{J.~S.}},
  \bibinfo{author}{\bibfnamefont{F.~B.~S.} \bibnamefont{{Paerels}}},
  \bibinfo{author}{\bibfnamefont{F.}~\bibnamefont{{Durret}}}, \emph{et~al.},
  \bibinfo{year}{2008}, \bibinfo{journal}{\ssr} \textbf{\bibinfo{volume}{134}},
  \bibinfo{pages}{155}.

\bibitem[{\citenamefont{{Mac{\'{\i}}as-Ram{\'{\i}}rez}}
  \emph{et~al.}(2012)\citenamefont{{Mac{\'{\i}}as-Ram{\'{\i}}rez}, {Gordon},
  {Brown}} \emph{et~al.}}]{Macias-Ramirez:2012aa}
\bibinfo{author}{\bibnamefont{{Mac{\'{\i}}as-Ram{\'{\i}}rez}},
  \bibfnamefont{O.}},
  \bibinfo{author}{\bibfnamefont{C.}~\bibnamefont{{Gordon}}},
  \bibinfo{author}{\bibfnamefont{A.~M.} \bibnamefont{{Brown}}}, \emph{et~al.},
  \bibinfo{year}{2012}, \bibinfo{journal}{\prd}
  \textbf{\bibinfo{volume}{86}}(\bibinfo{number}{7}), \bibinfo{eid}{076004}.

\bibitem[{\citenamefont{{Navarro}} \emph{et~al.}(1997)\citenamefont{{Navarro},
  {Frenk}, and {White}}}]{Navarro:1997aa}
\bibinfo{author}{\bibnamefont{{Navarro}}, \bibfnamefont{J.~F.}},
  \bibinfo{author}{\bibfnamefont{C.~S.} \bibnamefont{{Frenk}}}, and
  \bibinfo{author}{\bibfnamefont{S.~D.~M.} \bibnamefont{{White}}},
  \bibinfo{year}{1997}, \bibinfo{journal}{\apj} \textbf{\bibinfo{volume}{490}},
  \bibinfo{pages}{493}.

\bibitem[{\citenamefont{{Petrosian}}
  \emph{et~al.}(2008)\citenamefont{{Petrosian}, {Bykov}, and
  {Rephaeli}}}]{Petrosian:2008aa}
\bibinfo{author}{\bibnamefont{{Petrosian}}, \bibfnamefont{V.}},
  \bibinfo{author}{\bibfnamefont{A.}~\bibnamefont{{Bykov}}}, and
  \bibinfo{author}{\bibfnamefont{Y.}~\bibnamefont{{Rephaeli}}},
  \bibinfo{year}{2008}, \bibinfo{journal}{\ssr} \textbf{\bibinfo{volume}{134}},
  \bibinfo{pages}{191}.

\bibitem[{\citenamefont{{Pinzke} and {Pfrommer}}(2010)}]{Pinzke:2010aa}
\bibinfo{author}{\bibnamefont{{Pinzke}}, \bibfnamefont{A.}}, and
  \bibinfo{author}{\bibfnamefont{C.}~\bibnamefont{{Pfrommer}}},
  \bibinfo{year}{2010}, \bibinfo{journal}{\mnras}
  \textbf{\bibinfo{volume}{409}}, \bibinfo{pages}{449}.

\bibitem[{\citenamefont{{Pinzke}} \emph{et~al.}(2011)\citenamefont{{Pinzke},
  {Pfrommer}, and {Bergstr{\"o}m}}}]{Pinzke:2011aa}
\bibinfo{author}{\bibnamefont{{Pinzke}}, \bibfnamefont{A.}},
  \bibinfo{author}{\bibfnamefont{C.}~\bibnamefont{{Pfrommer}}}, and
  \bibinfo{author}{\bibfnamefont{L.}~\bibnamefont{{Bergstr{\"o}m}}},
  \bibinfo{year}{2011}, \bibinfo{journal}{\prd}
  \textbf{\bibinfo{volume}{84}}(\bibinfo{number}{12}), \bibinfo{eid}{123509}.

\bibitem[{\citenamefont{{S{\'a}nchez-Conde}}
  \emph{et~al.}(2011)\citenamefont{{S{\'a}nchez-Conde}, {Cannoni}, {Zandanel}}
  \emph{et~al.}}]{Sanchez-Conde:2011aa}
\bibinfo{author}{\bibnamefont{{S{\'a}nchez-Conde}}, \bibfnamefont{M.~A.}},
  \bibinfo{author}{\bibfnamefont{M.}~\bibnamefont{{Cannoni}}},
  \bibinfo{author}{\bibfnamefont{F.}~\bibnamefont{{Zandanel}}}, \emph{et~al.},
  \bibinfo{year}{2011}, \bibinfo{journal}{\jcap} \textbf{\bibinfo{volume}{12}},
  \bibinfo{eid}{011}.

\bibitem[{\citenamefont{{S{\'a}nchez-Conde} and
  {Prada}}(2014)}]{Sanchez-Conde:2014aa}
\bibinfo{author}{\bibnamefont{{S{\'a}nchez-Conde}}, \bibfnamefont{M.~A.}}, and
  \bibinfo{author}{\bibfnamefont{F.}~\bibnamefont{{Prada}}},
  \bibinfo{year}{2014}, \bibinfo{journal}{\mnras}
  \textbf{\bibinfo{volume}{442}}, \bibinfo{pages}{2271}.

\bibitem[{\citenamefont{{The Fermi-LAT
  Collaboration}}(2015)}]{The-Fermi-LAT-Collaboration:2015aa}
\bibinfo{author}{\bibnamefont{{The Fermi-LAT Collaboration}}},
  \bibinfo{year}{2015}, \bibinfo{journal}{ArXiv e-prints} \eprint{1501.02003}.

\bibitem[{\citenamefont{{Vazza} and {Br{\"u}ggen}}(2014)}]{Vazza:2014aa}
\bibinfo{author}{\bibnamefont{{Vazza}}, \bibfnamefont{F.}}, and
  \bibinfo{author}{\bibfnamefont{M.}~\bibnamefont{{Br{\"u}ggen}}},
  \bibinfo{year}{2014}, \bibinfo{journal}{\mnras}
  \textbf{\bibinfo{volume}{437}}, \bibinfo{pages}{2291}.

\end{thebibliography}


\end{document}